\documentclass[fleqn,usenatbib]{mnras}

\usepackage[T1]{fontenc}
\usepackage{ae,aecompl}

\usepackage{graphicx}	
\usepackage{amsmath}	
\usepackage{amssymb}	

\usepackage{multirow}

\usepackage{newtxtext,newtxmath}

\newcommand{\pc}{\,{\rm pc}}
\newcommand{\kms}{\,{\rm km}\,{\rm s}^{-1}}
\newcommand{\msun}{\, {\rm M_{\odot}}}
\newcommand{\rsun}{\, {\rm R_{\odot}}}

\title[Primordial black hole capture by stars]{Primordial black holes capture by stars and induced collapse to low-mass stellar black holes}

\author[M. Oncins et al.]{
Marc Oncins,$^{1,2,5}$ 
Jordi Miralda-Escud\'e,$^{1,2,3,5}$
Jordi L. Guti\'errez$^{4,5}$ and Pilar Gil-Pons$^{4,5}$
\\
$^{1}$Institut de Ciencies del Cosmos (ICCUB), Universitat de Barcelona (UB-IEEC), Mart\'i i Franqu\`es 1, 08028 Barcelona, Spain\\
$^{2}$Departament de F\'isica Qu\`antica i Astrof\'isica, Facultat de F\'isica, Universitat de Barcelona, 
Mart\'i i Franqu\`es 1, 08028 Barcelona, Spain \\
$^{3}$Instituci\'o Catalana de Recerca i Estudis Avan\c cats, Barcelona, Spain\\
$^{4}$EETAC, Universitat Polit\`ecnica de Catalunya, Campus Baix Llobregat, C3, 08840 Castelldefels, Spain. \\
$^{5}$Institut d'Estudis Espacials de Catalunya, Ed- Nexus Campus Nord, Barcelona, Spain.
}

\date{}

\pubyear{2022}

\begin{document}
\label{firstpage}
\pagerange{\pageref{firstpage}--\pageref{lastpage}}
\maketitle

\begin{abstract}
Primordial black holes in the asteroid-mass window, which might constitute all the dark matter, can be captured by stars when they traverse them at low enough velocity. After being placed on a bound orbit during star formation, they can repeatedly cross the star if the orbit happens to be highly eccentric, slow down by dynamical friction and end up in the stellar core. The rate of these captures is highest in halos of high dark matter density and low velocity dispersion, when the first stars form at redshift $z \sim 20$. We compute this capture rate for low-metallicity stars of $0.3$ to $1\msun$, and
find that a high fraction of these stars formed in the first dwarf galaxies would capture a primordial black hole, which would then grow by accretion up to a mass that may be close to the total star mass. We show the capture rate of primordial black holes does not depend on their mass over this asteroid-mass window, and should not be much affected by external tidal perturbations. These low-mass stellar black holes could be discovered today in low-metallicity, old binary systems in the Milky Way containing a surviving low-mass main-sequence star or a white dwarf, or via gravitational waves emitted in a merger with another compact object. No mechanisms in standard stellar evolution theory are known to form black holes below the Chandrasekhar mass, so detecting a low-mass black hole would fundamentally impact our understanding of stellar evolution, dark matter and the early Universe.
\end{abstract}




\section{Introduction}

Dark matter (DM) remains one of the most important mysteries in modern cosmology. One of the hypotheses for the nature of DM are Primordial black holes (PBHs) made in the early Universe from large amplitude primordial fluctuations \citep{1971MNRAS.152...75H, 1974MNRAS.168..399C, 1975Natur.253..251C, 1975ApJ...201....1C}. The abundance of PBHs is limited by several observational constraints;
reviews over the entire PBH mass range can be found, for example, in \citet{2018CQGra..35f3001S,2021RPPh...84k6902C,2021arXiv211002821C}. While their abundance is severely limited below a mass of $\sim 10^{-16.5}\msun$ by the contribution from Hawking evaporation to the $\gamma$-ray background and other impacts on the Cosmic Microwave Background \citep{2020PhRvD.101l3514L}, PBHs might still constitute all of the DM in the Universe in the asteroid-mass range from $10^{-16} \msun$ to $10^{-11}\msun$.
Despite several early claims for closing this mass window, follow-up studies raised objections on their validity \citep{2019JCAP...08..031M,2020PhRvD.101f3005S}. In addition, PBH might
account for a substantial fraction, but not all, the DM at higher masses while avoiding gravitational microlensing and other constraints \citep{2000ApJ...542..281A,2019NatAs...3..524N, 2020PhRvD.101f3005S}.

Alternative ways to find observational consequences of PBHs in this mass range are therefore of strong interest. A novel approach was proposed in \citet{2009arXiv0901.1093R}, where the idea that these asteroid-mass PBHs might randomly traverse through a star and be captured in the stellar core was presented. If the DM contains PBHs, these would be present in the cool molecular gas clouds where stars form and, if the relative velocities are low enough, they would adiabatically follow the gas contraction during the formation of a protostar, resulting in PBHs bound to the newly born star. A portion of the PBHs in highly eccentric orbits would lose energy via dynamical friction when crossing through the stellar interior and fall to the stellar core. Once settled in the star centre, a PBH would start accreting, growing up to a total mass that can be close to the total mass of the star.

This suggests the possibility to form a black hole with a typical mass of a star but below the Chandrasekhar mass, which cannot be explained in ordinary stellar evolution. Possible origins of these transmuted black holes, as they have been named \citep{2018PhLB..782...77T}, may be particle DM \citep{2018PhRvL.121v1102K, 2021PhRvL.126n1105D} and PBH capture by neutron stars \citep{2018ApJ...868...17A, 2020PhRvD.102h3004G}, but most require the high density of neutron stars. If a black hole with mass below the Chandrasekhar limit were discovered, it would strongly point to an origin in a PBH or some other process involving DM interaction with stars. However, \citet{2009ApJ...705..659A} questioned that this process could occur in the Milky Way, finding the rate of PBH capture in normal stars to be negligible for the present low DM density and high velocity dispersion.

The capture of PBH was also studied by \citet{2009MNRAS.399.1347B}, focusing on massive stars as the possible origin of supermassive black holes at high redshift. The higher density and lower velocity dispersion of DM halos in the early universe leads to more common PBH capture in the first metal-free stars, although these would not leave unique observational signatures, their mass being similar to other stellar black holes formed at the end of the lives of the same stars.

PBH capture in the present Universe was reexamined by \citet{2013PhRvD..87b3507C}, considering star formation in globular clusters formed in dense DM halos made of PBHs in the asteroid-mass range. The impact of eccentric orbits to the capture rate was included in \citet{2014PhRvD..90h3507C}, who found an enhanced capture rate implying that no neutron stars would form because all their progenitors would have captured a PBH that would accrete the star before or during the formation of the neutron star. 
This work, however, considered globular clusters with a very high DM and baryonic density, whereas in fact globular clusters may form from gas cloud fragmentation without involving any DM, so their constraints are not readily applicable \citep{2019JCAP...08..031M}.

In this paper we seek to expand on previous work by studying the effect of PBH capture in the evolution and fate of low-mass stars at high redshift, using precise stellar models of low-mass stars with a range of masses from $0.32\msun$ to $1\msun$, and with very low metallicity (as expected for the first stars). We consider the first stars formed in the Universe at $z \sim 20$, when both DM density is highest and the velocity dispersion in the star-forming DM halos is lowest, therefore maximising the capture rate of PBHs by stars. We focus on low-mass stars because they are uniquely able to produce black holes of stellar mass below the Chandrasekhar value, giving a clear signature that cannot be explained by standard stellar evolution theory. Although the first stars to form from metal-free gas are expected to have a top heavy initial mass function, stars of lower mass should form almost immediately thereafter from gas polluted by the first few supernovae, so they were probably abundantly made in a way similar to present-day galaxies.

The methods of our calculation are described in section \ref{sec:Methods}. In section \ref{sec:Results} we present our results for the PBH capture rates for stellar models of various stellar masses. Finally our conclusions are discussed in section \ref{sec:Conclusions}. We assume a flat cosmology with $\Omega_b = 0.3$ and $H_0= 70 \, \rm km s^{-1} Mpc^{-1}$

\section{Capture of black holes by a Star: Methods}\label{sec:Methods}

Our aim in this section is to calculate the rate at which PBH accounting for the DM in a halo where a star forms are captured by randomly traversing the star and being slowed down by dynamical friction.

\subsection{Dark Matter density profile around first stars}\label{sec:rhoPBH}

 According to the standard Cold Dark Matter model, the first stars
should form in the first halos where the gravitationally collapsed gas
that is shock-heated to the halo virial temperature is able to
radiatively cool. For the gas of primordial composition, this happens
first when trace amounts of molecular hydrogen formed from the remnant
ionization that is left over from the recombination epoch induce a
cooling rate that is higher than the characteristic inverse time between
successive halo mergers, at $z\simeq 20$ in halos of $M\sim 10^6 \msun$
\citep[e.g.][]{1984Natur.311..517B, 1997ApJ...474....1T}

Analytic studies and numerical simulations initially suggested that
stars formed in these halos from primordial gas are all of high mass
\citep[][]{2002Sci...295...93A, 2014ApJ...781...60H}. Newer results however have shown strong fragmentation is possible, resulting in zero metallicity low-mass star formation \citep{2002ApJ...569..549N,2011Sci...331.1040C,2015MNRAS.447.3892H, 2014ApJ...792...32S} and even binary systems \citep{2018MNRAS.479..667R}. Even ignoring this and assuming low-mass stars do not form at zero metallicity, they
should start forming soon thereafter. In this paper we are interested
in low-mass stars, which can survive to the present time either as red
dwarfs in the main-sequence or as white dwarfs, but may become a low-mass black hole if they have captured a PBH from the surrounding DM.

We consider that these low-mass stars acquired a quantity of bound DM at
birth owing to adiabatic contraction during the collapse of the gas
cloud that formed the star \citep{2013PhRvD..87b3507C}, which thereafter remains
bound to the star for an arbitrarily long time. In other words, we
assume that the original DM that is left bound to the star is not
removed by tidal disruption at a later time; we will discuss this further when considering the impact of external tides.

We take as fiducial values of the mass and formation time of the halo
that hosts the first generation of low-mass stars $M=10^7\msun$ and
$z=20$, to be conservative on the minimum halo mass where low-mass stars
start forming. The mass of DM left bound to the star
within a given radius depends only on the phase-space density of DM in the formation site. We define the virial halo density
$\rho_v$ and virial radius $R_v$ as
\begin{align}\label{eq:densh}
   \rho_v &= \rho_c \Delta_c = 18\pi^2 \rho_c \simeq 0.22
 \left( \frac{1+z}{21} \right)^3 \msun\pc^{-3} ~,  \\
   R_v &= \left( \frac{3M}{4\pi \rho_v} \right)^{1/3} \simeq
 220\, \left( \frac{M}{10^7 \msun} \right)^{1/3} \frac{21}{1+z} \pc ~,
   \label{eq:rho}
    \end{align}
where $\rho_c=3H^2/(8\pi G)$ is the critical density of the Universe
and we use the critical overdensity of the top-hat spherical model at
virialization, $\Delta_c=18\pi^2$. The implied halo velocity dispersion
depends of course on the assumed density profile, but is approximately
\begin{equation}\label{eq:sigmah}
 \sigma_v \simeq \left( \frac{GM}{2R_v}\right)^{1/2} \simeq 9.9
 \left( \frac{M}{10^7 \msun} \right)^{1/3}\,
 \left( \frac{1+z}{21} \right)^{1/2} \kms ~.
\end{equation}

The phase-space density of DM around the star depends on its distance
from the halo center at formation time, increasingly rapidly towards the
center. We assume the halo has a standard Navarro-Frenk-White (NFW,
\citet{1997ApJ...490..493N}) density profile, as found in numerical
simulations of structure formation. The profile is characterized by a
scale radius $R_s=R_v/c$, where $c$ is the concentration parameter:
\begin{align}
  \rho_h (R) = \frac{\rho_{0}}{R/R_s \left(1+R/R_s \right)^2} ~,
  \label{eq:rho_vir}
\end{align}
where R is the radial distance to the center of the halo and $\rho_0$ is a normalization constant that we determine by
requiring the total mass within $R_v$ to be $M$. We use the value
$c=10$ throughout this paper, in agreement with simulations of the
early halo formation \citep{2001MNRAS.321..155L}.

The detailed phase-space density of the DM at a given radius $R$
depends on the velocity as determined by the condition of dynamical
equilibrium. For simplicity, we assume a Gaussian velocity distribution
to compute the phase-space number density
for small velocities,
\begin{equation}
 Q_g= \frac{\rho_h(R)}{m_b \left[2\pi\sigma_h^2(R)\right]^{3/2} } ~,
\end{equation}
where $\sigma_h(R)$ is the DM one-dimensional velocity dispersion in
the halo as a function of radius, assuming an isotropic model, and $m_b$ is the mass of each PBH. The
DM phase-space density bound to the star is assumed to have a constant
value $Q=f_s Q_g$, where $f_s$ is a dimensionless constant of
order unity that depends on the velocity of the star, and is used also to absorb the difference between the detailed velocity distribution of the NFW profile at a specific radius $R$ and a Gaussian distribution. The maximum amount of bound
DM will be acquired when the star is at rest relative to the mean
surrounding DM, but typically the star is moving at an
rms velocity $\sim \sqrt{3}\sigma_h$ and will be acquiring bound DM at the
phase-space density near the star velocity.

The density profile of bound DM, $\rho_{bd}(r)$, after the star of mass
$M_*$ is formed, at a distance $r$ from the star, is determined by an
approximately constant DM phase-space density up to a maximum of the escape velocity
relative to the star $v=(2GM_*/r)^{1/2}$, and is therefore given by
\begin{equation}
\label{eq:rbd}
 \rho_{bd}= Q m_b \int_0^{\sqrt{2GM_*/r}} 4\pi v^2\, dv =
 \frac{4f_s}{3 \sqrt{\pi} }\, \frac{\rho_h(R)}{\sigma_h(R)^3}\,
 \left(\frac{GM_*}{r} \right)^{3/2} ~.
\end{equation}
Therefore the DM that is left bound around the star has a density
profile proportional to $r^{-3/2}$, with a normalization that is
determined by the density and velocity dispersion of the dark matter
halo at the formation site of the star, and the dimensionless factor of
order unity $f_s$. In this paper we will consider only the capture of
this bound DM, and neglect any unbound DM that may be also be captured
by the star when traversing it, if the incoming velocity is low enough
to allow capture after a single passage through the star. From \citet{2009ApJ...705..659A, 2013PhRvD..87b3507C} we expect the total capture rate should generally be dominated by this
originally bound DM, especially when taking into account the effects of likely
external perturbations on the DM orbits around the star that can
randomly change orbital eccentricities.

\begin{figure}
	\includegraphics[width=\columnwidth]{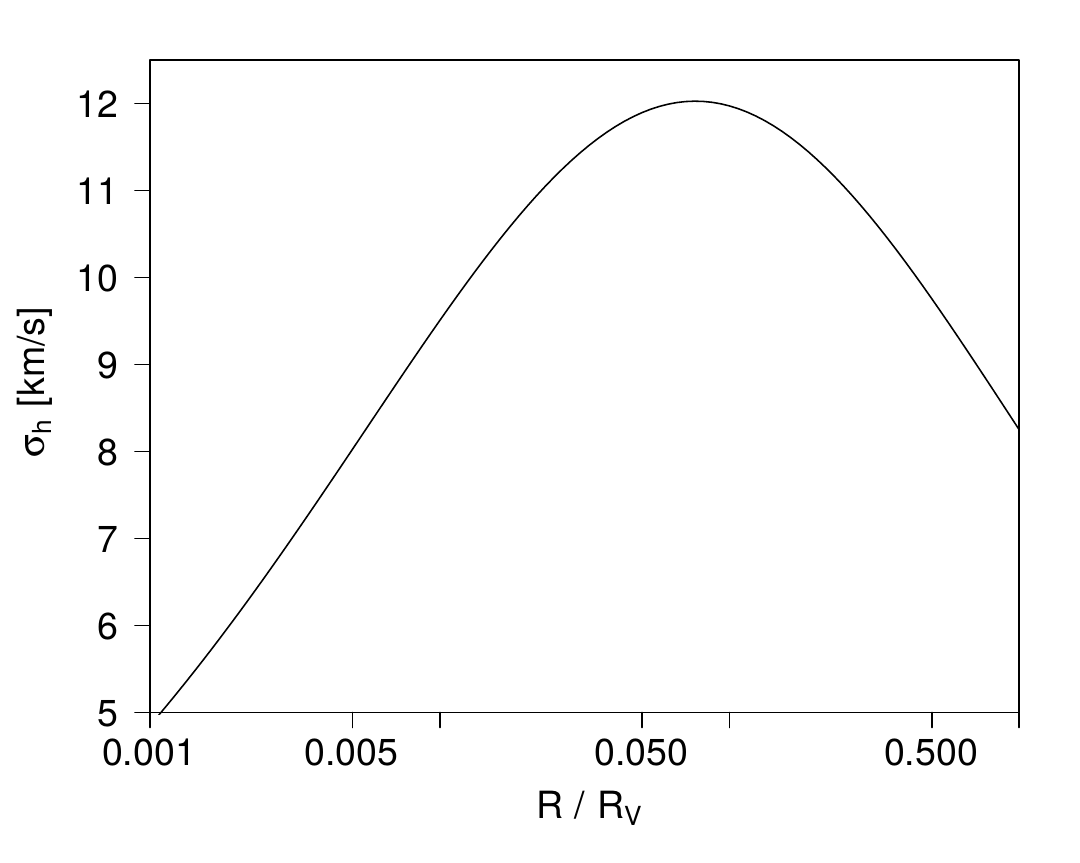}
    \caption{One-dimensional velocity dispersion of an isotropic NFW halo of mass $M=10^7 \msun$ at $z=20$, 
    used to compute the DM phase-space density around a star formed at a radius $R$ in the halo according to equation (5).}
    \label{fig:sigma_halo}
\end{figure}

We compute the DM phase-space density in our model using $\sigma_h(R)$ from equation (13) of \citet{2001MNRAS.321..155L}, for the isotropic case. For reference, we show in Figure \ref{fig:sigma_halo} the computed one-dimensional velocity dispersion $\sigma_h(R)$, for our fiducial model of a halo mass $M=10^7\msun$ at $z=20$.

\subsection{Black Hole capture: dynamical friction inside the star}\label{sec:PBH_capt}

Once the PBH is in a bound orbit around the star, the mechanism to dissipate its orbital energy and gradually fall to the core of the star is dynamical friction when it traverses the stellar interior. This requires the orbital eccentricity to be close enough to one for the periastron to be in the stellar interior.

To compute the dynamical friction effect accurately, a stellar model for the interior density, temperature, and sound speed profiles must be used, which will be described below in section \ref{sec:Stellar}.
The orbit external to the star is purely Keplerian, characterized by the specific energy $u$ and angular momentum $\ell$,

\begin{align}
    \ell = 
    \sqrt{GM_{\ast} a \left(1-e^2\right)} ~,
    \label{eq:mom}
\end{align}
\begin{align}
    u = \frac{-G M_{\ast}}{2a} = \frac{1}{2}v^2 - \frac{GM_*}{r} \;,
    \label{eq:ener}
\end{align}
where $e$ is the eccentricity and $a$ the semimajor axis.

To compute the PBH capture rate, we need to calculate the time required for capture as a function of the semimajor axis and orbital eccentricity. This is determined by the amount of energy the PBH loses per stellar crossing. As long as the PBH periastron is within the stellar radius, the energy dissipation process continues with a decreasing semimajor axis and orbital period, and more frequent crossings that result in faster energy dissipation.

Therefore, to compute the time required for capture we do not need to worry about the detailed evolution of the energy and angular momentum of the orbit as the PBH is gradually slowed down through a large number of stellar crossings, because the semimajor axis remains much larger than the periastron, implying a negligible change in the part of the orbit in the stellar interior,
over most of the capture process time. In fact, we can simply compute the loss of energy over one stellar crossing by integrating over the unperturbed orbit of the PBH through the interior of the star, because the change in orbital energy is very small in a single stellar crossing compared to the kinetic energy of the PBH in the stellar interior.

We also approximate dynamical friction to depend only on the local plasma density and sound speed in the PBH vicinity (this is equivalent to neglecting the contribution to the Coulomb logarithm from the largest distances, comparable to the stellar radius, where these thermodynamic variables have substantial variation). Then, the dynamical friction acceleration $a_{df}$ is always in the direction opposite to the orbital velocity $v$, and the rate of energy loss is $du= - a_{df}\, v\, dt$, so the loss of specific orbital energy per stellar crossing is given by
\begin{equation}
  \Delta u= - \int_{t1}^{t2} a_{df}\, v \, d t = - 2\int_q^{R_{\ast}} a_{df} \, v \frac{d t}{d r}\, dr ~,
\end{equation}
where the first integral is from the time $t_1$ when the PBH enters the star to the time $t_2$ when it exits it. In the second integral, we change variables to the radial coordinate $r$ and use the fact that the integral contains two symmetric parts, from the stellar radius $R_*$ to the periastron $q$ and viceversa. 

As explained before, we compute this integral for the unperturbed orbit of the PBH moving in the gravitational field of the star without the dynamical friction, because the modification of the orbit in a single passage is very small. The radial time derivative is related to the conserved specific angular momentum as $dr/dt=v[1-(\ell/rv)^2]^{1/2}$, so the above integral becomes
\begin{equation}\label{eq:Du}
  \Delta u = - 2\int_q^{R_{\ast}} \frac{a_{df}}{\sqrt{1-(\ell/rv)^2}}\,  dr ~.
\end{equation}

The acceleration caused by dynamical friction is the quantity that depends on the stellar interior model. A first approximation one can use is
Chandrasekhar's formula \citep[][]{1949RvMP...21..383C}, but this is valid for collisionless matter only, which does not apply to the interior plasma in stars. Instead, the adequate computation to use is for a collisional fluid, as presented by \citet[][]{1999ApJ...513..252O}. This fluid friction is close to the collisionless formula for a Mach number $M_a=v/c_s > 2$, where $c_s$ is the sound speed, but is substantially larger  for $1<M_a<2$, a common value for the capture process because the free-fall speed of the PBH is larger than the sound speed in the stellar interior by a small factor (as an example, the escape velocity at the surface of the Sun is $v\approx 615\, {\rm km/s}$ compared to a typical sound speed of $c_s \sim 350 \, {\rm km/s}$ \citep{2001ApJ...555..990B}). For the collisional case, we will actually use a formula from a simplified analytic estimate by \citet{2016A&A...589A..10T}, obtained from numerical simulations.

In general, the dynamical friction can be written as
\begin{align}\label{eq:adf}
  a_{df} = \frac{4\pi G^2 \rho_s m_b}{v^2} \, I(v,\Lambda) ~,
  \end{align}
where $\rho_s$ is the density of the star, $m_b$ is the PBH mass, and the dimensionless function $I(v,\Lambda)$ contains the detailed physical dependence on the velocity dispersion and Coulomb logarithm for the collisionless or collisional cases. The Chandrasekhar formula is written as
  \begin{align}
 I_C(v, \Lambda) = \left[ \, {\rm erf} \left( \frac{v}{\sqrt{2}\sigma_s}\right) -\frac{2 v}{\sqrt{2\pi}\sigma_s}\, \exp\left(-\frac{v^2}{2\sigma_s^2} \right) \right]\, \ln\Lambda ~,
    \label{eq:Chandra}
  \end{align}
while for the fluid case, the equation from \citet{2016A&A...589A..10T} is
  \begin{align}
 I_T(v, \Lambda) = {\rm ln} \left[ 2\Lambda \left(1-1/M_a^2 \right) \right] ~,
    \label{eq:Thun}
  \end{align}
where erf is the error function, $\sigma_s = (3kT/\mu)^{1/2}$ is the three-dimensional velocity dispersion for the plasma, and $\Lambda = R_{\rm max}/R_{\rm min}$ is the Coulomb logarithm. The lengths $R_{\rm max}$ and $R_{\rm min}$ are the usual maximum and minimum impact parameters for an effective gravitational interaction to produce dynamical friction, while $T$ is the temperature, $k$ the Boltzmann constant, and $\mu$ the mean particle mass. Equation (\ref{eq:Thun}) is valid for $M_a>1$, which is generally correct for an object moving near the escape speed in the stellar interior. In addition, the factor of $2$ in this equation is a numerical result obtained for an adiabatic index of $5/3$ for the plasma, the value for a monatomic gas that we assume here.
We use the radius of the star for $R_{\rm max}$, and $R_{min}=(2Gm_b)/v^2$. Below $R_{\rm min}$, the deflection of gas by the PBH gravity is much less effective at slowing it down; in particular, any accretion into the PBH is negligible as far as the rate of slowing down its velocity is concerned. 
  
 \subsection{Number of Captured Black Holes}\label{sec:PBH_num}
  
If a PBH follows the trajectory determined only by the gravity of the star and the dynamical friction when it traverses the stellar interior calculated in the previous subsection, it will slow down gradually and reduce its semimajor axis until the orbit moves entirely to the interior of the star and the PBH settles on the stellar core. For the PBH to complete the process of orbital energy loss during the present age of the Universe, the initial orbital period has to be short enough so that the energy loss at each passage can add up roughly to the initial orbital energy. 

As in previous work \citep[e.g.,][]{2013PhRvD..87b3507C}, 
we assume the orbital energy loss at each crossing of the stellar interior, $| \Delta u|$, is small compared to $| u|$, so we can integrate the evolution of the orbital energy with time $t$ with the simple equation
  \begin{equation}
      du = \Delta u \, \frac{dt}{P} = \Delta u \frac{dt}{P_0} \left( \frac{u}{u_0} \right)^{3/2} ~, 
  \end{equation}
where $u_0$ is the initial orbital energy and $P_0=\pi GM_* (|u_0|)^{-1}\, (2|u_0|)^{-1/2}$ is the initial orbital period. Solving this equation, we find the time $t_c$ required to capture the PBH to the stellar interior (i.e., to reduce the semimajor axis to a value much smaller than the initial one) is
\begin{equation}\label{eq:tc}
    t_c = P_0\, \frac{2u_0}{\Delta u} ~.
\end{equation}
Note that as long as the semimajor axis remains much larger than the stellar radius, the trajectory of the PBH through the stellar interior at each crossing remains nearly the same, so $\Delta u$ stays almost constant and this simple solution is a good estimate. The detailed orbital evolution during the late stages of the capture, when the semimajor axis becomes comparable to the stellar radius, do not matter because the orbital period is then very short and capture takes a short time to be completed. Most of the required time for capture is at large semimajor axis, close to its initial value. In our model we still compute the time and energy loss for each orbit allowing for $\Delta u$ to vary until the orbit is wholly within the star, but the analytic estimate gives results very close to the ones we found.

For each initial orbital energy $u_0$, there is a critical value of the eccentricity $e_c(u_0)$ that makes the capture time $t_c$ equal to the present age of the Universe. The condition for the PBH to be captured (in the absence of other orbital perturbations, which we will discuss below) is then that for a fixed energy the orbital eccentricity is larger than this critical value. We assume the distribution of eccentricities follows the thermal distribution that is implied when the phase-space density is constant, so the probability for the eccentricity to be above $e_c$ is $1-e_c^2$.

We can now calculate the total number of PBH that will be captured by the star over a time $t_c$. At a fixed radius $r$ from the star, the number density of PBH with phase-space density $Q$ that will be found with an orbital semimajor axis $a$, and therefore orbital energy $u=-GM_*/(2a)$ and velocity $v=(2GM_*/r-GM_*/a)^{1/2}$, is
\begin{equation}
    n_b(a)\, da= 4\pi Q v^2\, dv = {2\pi Q} \,(GM_*)^{3/2}\,
    \left(\frac{2}{r} - \frac{1}{a} \right)^{1/2}\, \frac{da}{a^2} ~,
\end{equation}
where we have used $v\, dv= GM_*\, da/(2a^2)$ at fixed $r$. Integrating over the volume, from $r=0$ to $r=2a$ and replacing $x=r/a$, the total number of PBH with semimajor axis $a$ is found to be
\begin{equation}
    N_b(a)\, da= {8\pi^2 Q}\, (GM_*)^{3/2}\, a^{1/2}\, da\, \int_0^2 dx\, x^2\,
    \sqrt{\frac{2}{x}-1} ~,
\end{equation}
which yields
\begin{equation}
    N_b(a)\, da= {4\pi^3 Q}\, (GM_*)^{3/2}\, a^{1/2}\, da ~.
\end{equation}
Finally, the total number of captured PBH is expressed as
\begin{equation}\label{eq:ncap}
    N_c = \frac{\sqrt{2\pi^3}\, f_s \rho_h}{m_b\sigma_h^3} (GM_*)^{3/2} \int_0^{a_m} \left[1-e_c^2(a)\right] \sqrt{a}\, da ~,
\end{equation}
where $a_m$ is the maximum semimajor axis allowing capture within the age of the universe, at which $e_c(a_m)=1$, and we have used equation (\ref{eq:rbd}) to express the phase-space density in terms of the DM density and velocity dispersion around the star at its formation time.

The critical eccentricity is related to a critical extrapolated periastron, $q_c(a)=a[1-e_c(a)]$, where the true periastron is larger than $q_c(a)$ because of the reduced gravitational potential in the stellar interior compared to the Kepler one, owing to the extended mass distribution. Typically, $q_c(a)$ is of order the radius of the stellar core, where the stellar density is close to the maximum, at the values of $a$ close to $a_m$ that dominate the contribution to the integral in equation (\ref{eq:ncap}). We define the effective mean capture periastron, $\bar q_c$, as
\begin{equation}\label{eq:barqc}
    \bar q_c\, 2\sqrt{a_m} = \int_0^{a_m} q_c(a)\,
    \frac{da}{\sqrt{a}} ~.
\end{equation}
The captured number of black holes is then, approximating $1-e_c^2\simeq 2(1-e_c)$,
\begin{equation}
    N_c= \frac{2(2\pi GM_*)^{3/2} f_s \rho_h}{m_b \sigma_h^3}\, \bar q_c \sqrt{a_m} ~.
    \label{eq:Capture_rate}
\end{equation}
It is also useful to express this in terms of fiducial values,
\begin{equation}
    N_c= \frac{f_s \rho_h/m_b}{10^{14} {\rm pc}^{-3} }\, 
    \left( \frac{M_{\ast}}{\msun} \right)^{3/2} \left(\frac{10\, {\rm km/s}}{ \sigma_h}\right)^3 \frac{\bar q_c}{0.05 R_{\odot}} \sqrt{ \frac{a_m}{\rm pc}} ~.
    \label{eq:generalized_capture_rate}
\end{equation}
From equations (\ref{eq:densh}), (\ref{eq:sigmah}) and (\ref{eq:rho_vir}), we find that for a star formed at radius $R\sim 0.1 R_v$, a typical density $\rho_h\sim 10 \msun\pc^{-3}$ is expected, so for PBHs of mass $m_b=10^{-12}\msun$ that are the DM and follow a thermal eccentricity distribution, a star
of mass close to $\msun$ would have a probability of order 0.1 to capture a black hole during the age of the Universe if $a_m$ is as large as a parsec.

It is useful to estimate at this point a rough value for $a_m$ for some fiducial parameters. The change in orbital energy per stellar crossing can be approximated, from equations (\ref{eq:Du}) and (\ref{eq:adf}), as
\begin{equation}
    \Delta u \sim \frac{8\pi G^2\rho_c r_c m_b}{v_e^2} \ln \Lambda ~,
\end{equation}
where the stellar density near the core is $\rho_c\sim 100\, {\rm g\, cm}^{-3}$, the stellar core has size $r_c\sim 10^{10}\rm cm$, $v_e\sim 1000 \rm km/s$ is the escape velocity from the core, and $\Lambda=R_{\rm max}/R_{\rm min}\simeq R_*v_e^2/(2Gm_b) \sim M_*/m_b$, so we find
\begin{align}\label{eq:anest}
    \Delta u\sim ~ 6\times 10^{-5} \frac{\rho_c r_c}{ 10^{12}{\rm g\, cm^{-2}} }\left(\frac{10^3\, {\rm km/s}}{ v_e}\right)^2  \frac{m_b}{10^{-12}\msun}\, \frac{{\rm km}^2}{{\rm s}^2 } ~.
\end{align}
Precise calculations of $\Delta u$ will be presented in Section \ref{sec:Results} for specific stellar models. For a capture time $t_c= 10^{10}\, {\rm yr}$ and $M_*=\msun$, the maximum semimajor axis at which equation (\ref{eq:tc}) is obeyed is
\begin{align}\label{eq:ams}
    a_m\sim (2\, {\rm pc})\,
    \left( \frac{\rho_c r_c}{10^{12}{\rm g\, cm^{-2}}} \frac{m_b}{10^{-12}\msun}
    \right)^2 \left(\frac{10^3\,{\rm km/s}}{v_e}\right)^4 ~.
\end{align}
However, the ideal case of a Keplerian orbit around the single star of mass $M_*$ is not realistic for the large values of $a_m$ implied for the typical parameters in equation (\ref{eq:ams}), because tidal perturbations by the host DM halo and possibly other factors perturb the orbit, as we discuss next. 
 
\subsection{Impact of Tidal Perturbations on the Capture Rate}\label{sec:Perturbations}

We have so far assumed that the PBH moves in a Kepler orbit around the star of mass $M_*$ with no gravitational perturbations. This assumption is clearly unrealistic for a PHB mass as low as $m_b\sim 10^{-12}\msun$, because the tidal acceleration caused by the host DM halo, $g_h$, at the maximum semimajor axis $a_m$ is
\begin{equation}
    g_h\sim \frac{GM_h a_m}{R^3} \sim g_* \frac{M_h a_m^3}{M_* R^3}~,
\end{equation}
where $g_*=GM_*/a_m^2$ is the gravitational acceleration from the star on the PBH. For $M_h/M_*\sim 10^6$, the external tidal acceleration is larger than $g_*$ at $a_m> 0.01 R$. Taking as an example a typical halo radius where the star is located as $R\sim 0.1 R_v\sim 10\, {\rm pc}$, we would expect any dark matter further than $\sim 0.1$ pc from the star to actually be tidally disrupted from the host halo tide.

Moreover, the external tide perturbs the orbital eccentricity of any PBH, deviating it from the nearly radial orbit required to cross the stellar interior. The change in orbital eccentricity over one period induced by the external tide is related to the change in specific angular momentum as
\begin{equation}
    \delta e\simeq \frac{\sqrt{1-e^2} \delta\ell}{e \sqrt{GM_* a} } ~.
\end{equation}
We can reasonably assume that the external perturbation causes a change $\delta \ell / \sqrt{GM_* a} \sim g_h/g_* $ over one period, so for nearly radial orbits ($1-e \ll 1$), the eccentricity perturbation in one orbit is
\begin{equation}\label{eq:deg}
    \delta e \sim \sqrt{2(1-e)}\, \frac{g_h}{g_*} ~.
\end{equation}
At the same time, the minimum eccentricity at each semimajor axis $a$ required for the PBH to be effectively slowed down as it crosses the stellar interior is $1-e_c(a)=\bar{q}_c(a)/a$. This small window of eccentricity was denominated the ``loss-cone'' in \citet{1976MNRAS.176..633F} to refer to the region in velocity space where an orbiting object is lost because of the interaction with the central object, but we use the term ``loss-cylinder'' here because of the cylinder shape of this velocity space region. When $\delta e > 1-e_c$, the PBH is removed from the loss-cylinder. This clearly has the effect of decreasing the capture rate at large semimajor axis.

However, at small semimajor axis the PBH capture rate can be increased by the external perturbations. The reason is that in the absence of perturbations, PBHs that are initially at $a\ll a_m$ and within the loss-cylinder are captured over a time $t\ll t_c$, so the loss-cylinder is depleted and further captures can occur only from PBH near the edges of the loss-cylinder that cross the star through the low-density envelope, with reduced friction and energy loss. When perturbations are present, the loss-cylinder is refilled and the capture rate increases back to its most effective rate.

For orbits of semimajor axis $a$, the ideal orbital perturbation rate that leads to the maximum PBH capture rate is that which produces, over a time $t_c$, a change in eccentricity of
\begin{equation}\label{eq:De}
    \Delta e \sim \frac{\bar q_c a_m^{1/2}}{a^{3/2} } ~,
\end{equation}
because the time required for the PBH to lose its orbital energy at $a$ when the periastron is within $\bar q_c$ is only $t_c\, (a/a_m)^{1/2}$, so the PBH will be captured if perturbations induce a random-walk of the eccentricity within the characteristic interval $\Delta e$ over time $t_c$.

In general, orbital perturbations can greatly reduce PBH captures from $a\sim a_m$ because PBH are always removed from the loss-cylinder before they are captured, but as $a$ is decreased, the interval $\delta e$ over which the eccentricity random-walks over time $t_c$ decreases until it equals $\Delta e$ in equation (\ref{eq:De}). At this value of $a$, the PBH capture rate will be roughly the same as it was at the maximum semimajor axis $a_m$ in the absence of perturbations. Therefore, we conclude that despite the presence of orbital perturbations, the PBH capture rate will always be roughly the same as obtained from equation (\ref{eq:ncap}). Perturbations imply that most PBH are actually captured from orbits much closer to the star than $a_m$, but the total capture rate should not be greatly modified.

It is possible that the bound DM around the star has been tidally stripped at some time down to a semimajor axis $a\ll a_m$, and subsequently any external perturbers are removed so that perturbations are absent but the PBH are no longer available at large $a$. In this case  $a_m\sim 0.1 pc$, so the capture rate may obviously be greatly reduced. As we expect the capture rate to have a dependence of $a_m^{1/2}$ at larger $a$, when the change in $\bar{q}_c$ becomes negligible, this should result in a capture rate less than one order of magnitude smaller compared to previous $a_m\sim 1 \, {\rm pc}$. We will assume here that enough DM has been retained to maintain the capture rate close to the value computed with equation (\ref{eq:ncap}) in the presence of tidal perturbations by refilling of the loss-cylinder.

\subsection{Black Hole Growth after Capture}

Once the PBH has settled in the center of the stellar core by the continuous action of dynamical friction, it will start growing in mass by accreting the surrounding stellar plasma. As discussed previously by \citet{
1995MNRAS.277...25M,
2009arXiv0901.1093R,
2009MNRAS.399.1347B,
2019JCAP...08..031M}, the PBH can accrete rapidly at the Bondi accretion rate if photons are trapped with the accreting plasma, and proceeds more slowly if an accretion disk is formed that can radiate efficiently and slow down the accretion by emitting close to the Eddington luminosity. The Bondi accretion rate is 
\begin{align}
\frac{\Dot{M}_{\rm B}}{m_b } = \, & \frac{\pi G^2 m_b \rho_s}{c_{s}^3} \simeq
 3.24 \cdot 10^{-6} \, {\rm yr}^{-1} ~ \times \nonumber \\
 & \frac{m_b}{10^{-12}\msun} \frac{\rho_s}{100 \, \rm g/cm^3} \left(\frac{c_{s}}{300 \rm km/s}\right)^{-3} ~,
 \end{align}
 where $\rho_s$ and $c_s$ are the plasma density and sound speed in the stellar center. We see that at the Bondi accretion rate and typical values presented above, even an initial PBH mass as low as $10^{-16}\msun$ will grow its mass in less than $10^{10}$ years.
 The growth rate becomes faster as the mass increases so the PBH can accrete all the stellar mass if Bondi accretion continues. If accretion becomes Eddington-limited at some stage due to the angular momentum of accreting matter and formation of an accretion disk, the accretion rate becomes 
 \begin{align}
\frac{\Dot{M}_{E}}{m_b} = \frac{4\pi G m_{p}}{c\;\eta\;\sigma_{T}} \simeq  2.2\cdot 10^{-8} \frac{0.1}{\eta} \rm yr^{-1} \, ,
 \end{align}
where $\rm m_{p}$ is the proton mass, $\eta$ is the radiative efficiency of the accretion disk matter and $ \sigma_{T}$ the Thompson cross-section. In this case, the constant e-folding time for mass growth is also much less than the present age of first stars $t_c\simeq 10^{10}$ years. Therefore, the PBH will continue to grow until it has accreted a substantial mass of the star.

If Bondi accretion continues, the star would only start being dynamically affected by the black hole accretion a few days before being completely accreted \citep{1995MNRAS.277...25M}. On the other hand, if an Eddington luminosity is emitted the total stellar luminosity will obviously be dominated by the accretion already when $m_b \sim 10^{-5}\msun$ for $M_*=1\msun$. At this late stage, formation of a jet or other mechanical energy release resulting from accretion may result in the ejection of the stellar envelope, and the end of PBH accretion, but the details of this final process are complex and it is not clear what the final mass of the PBH will be. Depending on stellar rotation and perhaps other stellar properties, the final mass may be close to the initial mass of the star, or may be much lower if mechanical energy can disperse the stellar material. 

\subsection{Stellar Models} \label{sec:Stellar}

In the present work, we consider low-mass stars formed at high redshift in low-mass DM halos. Low-mass stars live for a long time ($\gtrsim 10^{10}$ yrs) and thus might have a considerably high probability of capturing a PBH. Consequently, they might form low-mass black holes that, if discovered, would point to a formation process beyond the standard stellar formation channels.
Low-mass stars were expected to form soon after the first metal-free stars, once the first supernovae increased the heavy element abundance of the gas enough to lead to fragmentation and collapse of low-mass cloud cores
\citep{2021MNRAS.508.4767S, 2022MNRAS.510.4019P}. The concept of critical metallicity, that is, a minimum metallicity below which low-mass stars could not form \citep{2005IAUS..228..121B} seemed to have observational support \citep{2007MNRAS.380L..40F}. However recent observations of metal-poor stars in the halo keep pushing this low threshold to increasingly low values. The star by \citet{2014Natur.506..463K} holds the current record, with a metallicity [Fe/H]=-7.1 .

Even in the primordial metal-free gas, disks around massive stars might fragment into low-mass objects that form stars in the range $0.1$ to $1 \msun$ \citep{2002ApJ...569..549N, 2020ApJ...901...16D, 2011Sci...331.1040C,2015MNRAS.447.3892H, 2014ApJ...792...32S, 2022ApJ...925...28L}, though the Initial Mass Function (IMF) of the most primitive stars is still largely uncertain\citep{2016MNRAS.462.1307S, 2015MNRAS.447.3892H}. Any such stars formed near the center of a low-mass DM halo are the best candidates for PHB capture.

In this study, we use models of very metal poor stars,  that is, of $Z =10^{-4}$. This is the typical metallicity at which we expect the transition from a top-heavy to a bottom-heavy IMF to happen \citep{2013MNRAS.432L..46S,2019ffbh.book...67K, 2021MNRAS.503.2014S}. Six different stellar models with stellar masses in the range $0.32\msun < M_* < 1 \msun$, were computed for the present study with the open-source software instrument Modules for Experiments in Stellar Astrophysics (MESA) \citep{2011ApJS..192....3P, 2013ApJS..208....4P, 2015ApJS..220...15P, 2018ApJS..234...34P, 2019ApJS..243...10P}.
These models provide the stellar density, temperature, sound speed and mean atomic weight as functions of the radii for models ranging from the zero age main-sequence till times above the age of the universe, from which the loss of energy per orbit depending on orbital energy and angular momentum can be calculated as described in the previous subsections.

\begin{figure}
  \includegraphics[width=\columnwidth]{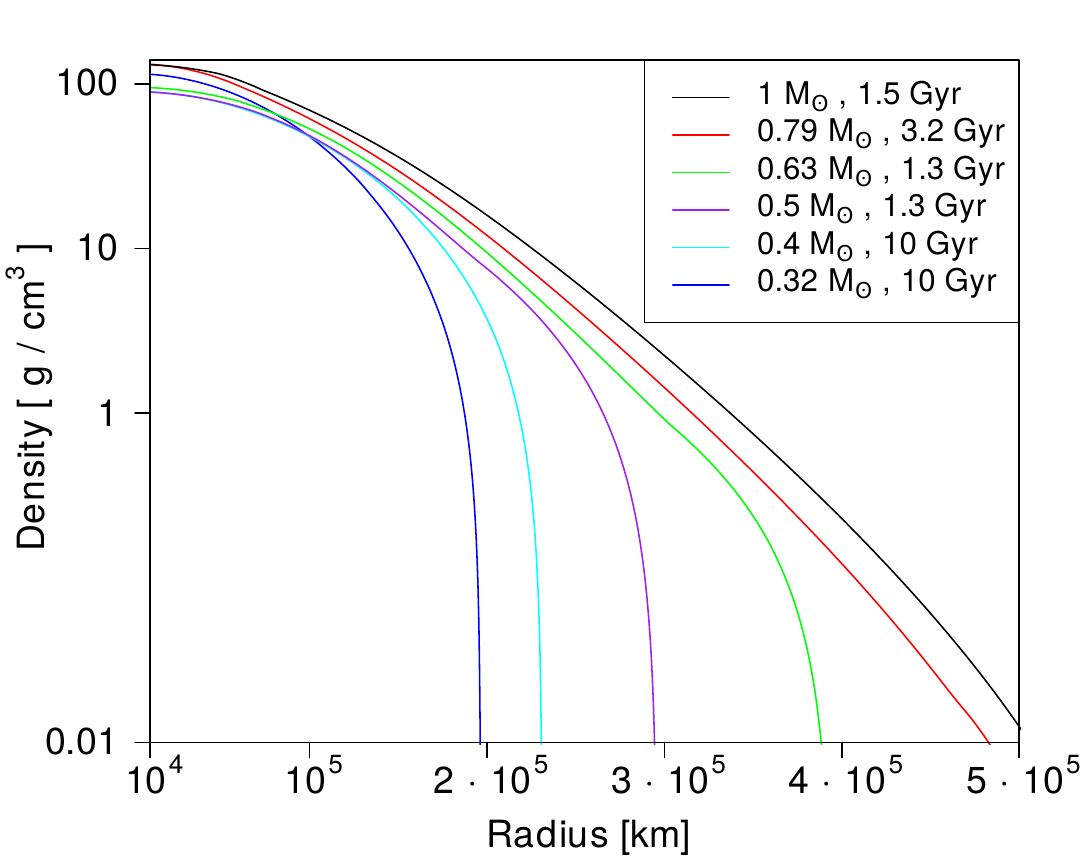}
  \caption{Mass density profile for the six stellar models used in this paper, with stellar masses 1, 0.79, 0.63, 0.5, 0.4, and 0.32 $\msun$, and ages 1.5, 3.2, 1.3, 1.3, 10, and 10 Gyr,  respectively.}
  \label{fig:Densstellmod}
\end{figure}

\begin{figure}
  \includegraphics[width=\columnwidth]{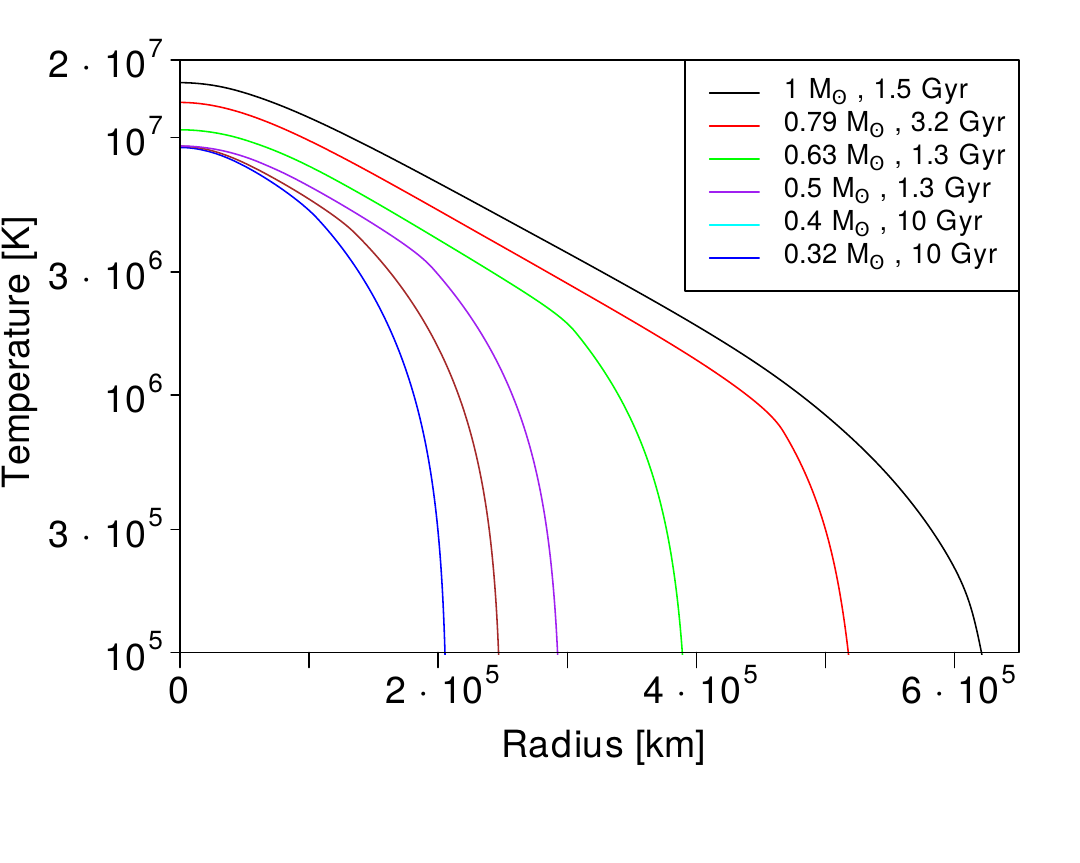}
  \caption{Temperature profiles for the same stellar models as in Figure \ref{fig:Densstellmod}.}
  \label{fig:Tempstellmod}
\end{figure}

A precise calculation of the PBH capture rate for a given stellar mass would involve averaging the capture rate over all ages, from the stellar birth to the present time. Instead of this, we present results for six cases of fixed stellar mass and age, assuming a constant stellar profile over the time $t_c$ at a fixed age. In practice, stars are in the main-sequence most of the time and low-mass stellar evolution is very slow, so this is a good approximation as long as we use an age when the star has already settled on its main-sequence. However, for stars of $M< 0.4 \msun$ the time required to stabilize near the main-sequence is as long as $\sim 10^9$ years, which causes a complex dependence of the central stellar density on mass if an early age is used. To illustrate the dependence on both mass and age, we present results for the six stellar models with masses
$1$, $0.79$, $0.63$, $0.5$, $0.4$, and $0.32 \msun$, and ages $1.5$, $3.2$, $1.3$, $1.3$, $10$, and $10$ Gyr, respectively. The density and temperature radial profiles of these six models are presented in Figures \ref{fig:Densstellmod} and \ref{fig:Tempstellmod}.


\section{Results for six stellar models}\label{sec:Results}

We now present specific results for our six stellar models, starting with the energy loss in a single stellar crossing, then the critical eccentricity for a given capture time of $t_c= 10^{10}$ years, and then the total number of captured black holes. Our results are presented for $m_b=10^{-12}\msun$, but can be scaled to other PBH masses in the asteroid mass range as good estimates in the way described previously.
 
\subsection{Energy loss by dynamical friction} \label{sec:Dyn_fric_rlts}
\begin{figure}
  \includegraphics[width=\columnwidth]{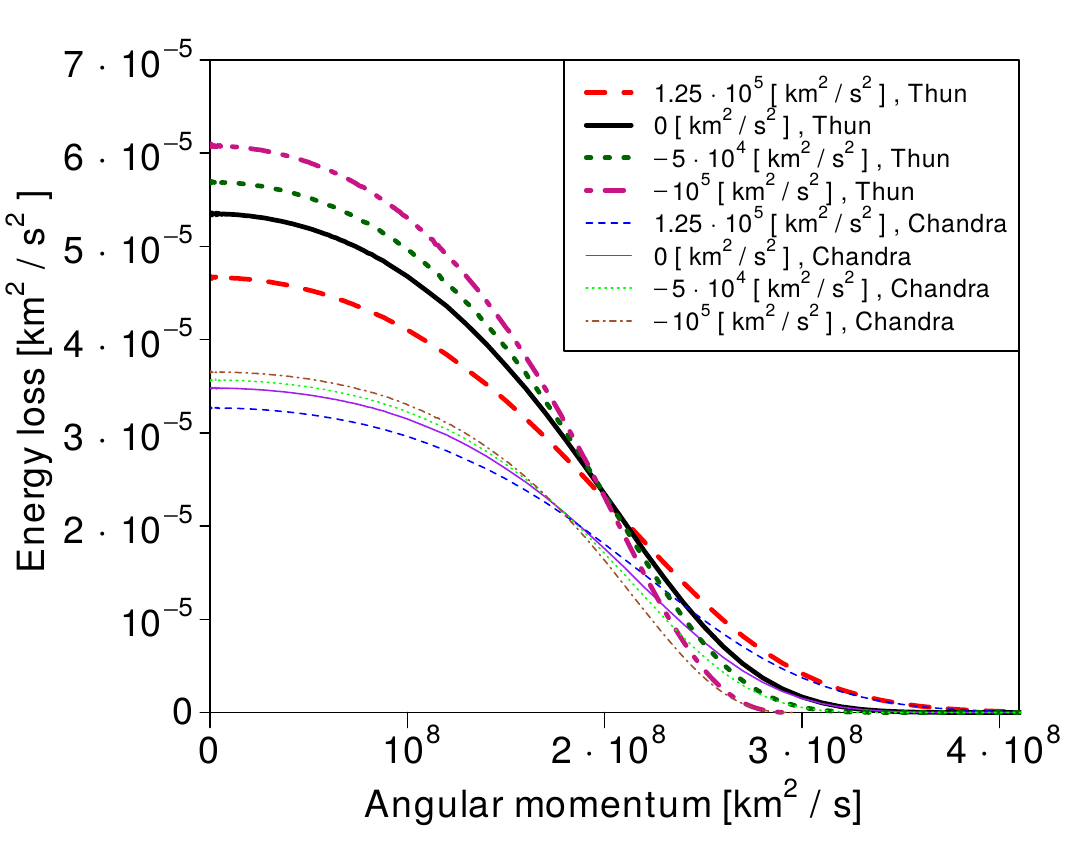}
  \caption{Specific energy loss of a $10^{-12}\msun$ PBH when crossing our model star with $M_*=1\msun$ and $t=1.5$ Gyr, as a function of the orbital
  angular momentum, using both the Chandrasekhar dynamical friction formula for collisionless matter (Chandra, thin line) and the hydrodynamic formula (Thun, thick line) based on \citet{2016A&A...589A..10T} and \citet{1999ApJ...513..252O}. The four curves of each case are for the indicated specific orbital energies (zero energy is a parabolic orbit, and negative energy is for bound elliptical orbits).}
  \label{fig:Eloss}
\end{figure}

Using equations (\ref{eq:Du}) and (\ref{eq:adf}), we can find the loss of energy over a single passage as a function of the PBH specific energy and angular momentum. As an illustration, we show this energy loss for the highest mass stellar model, with $M=1\msun$ and $t=1.5$ Gyr.
Results are plotted as a function of specific angular momentum in 
Figure \ref{fig:Eloss} for the collisionless case using equation (\ref{eq:Chandra}), and for the fluid case, using equation
(\ref{eq:Thun}), for four different values of the specific energy: the parabolic case with zero energy, an unbound case, and two bound orbits, with specific energies as indicated in the figures. The assumed PBH mass is $m_b=10^{-12}\msun$, but note that the energy loss is proportional to $m_b$, from equation (\ref{eq:adf}).

The energy loss $\Delta u$ in Figure \ref{fig:Eloss} agrees with the simple analytic estimate from equation (\ref{eq:anest}). When the PBH is captured from a semimajor axis much larger than the stellar radius, the relevant result for calculating the time it takes for the PBH to finalize the capture process is the one for the parabolic orbit, with zero total energy. The unbound case is never used for our computation (we neglect any contribution to the capture rate from unbound, incoming PBHs that were not placed in bound orbits when the star formed). The bound orbits show appreciable differences in $\Delta u$ in Figure \ref{fig:Eloss} only when the apoastron is already not much larger than the stellar radius; by then, the capture process is almost finished. The critical eccentricity $e_c(a)$ therefore depends, to an excellent approximation, on the zero energy curve for $a\gg R_*$. Note that the characteristic specific angular momentum where $\Delta u$ drops corresponds to $\sim v_e \bar q_c$, as we would expect from equation (\ref{eq:barqc}).

The hydrodynamical expression leads to an energy loss that is a factor $\sim 1.6$ larger than that obtained from the Chandrasekhar formula.

\subsection{Critical eccentricity for capture}

\begin{figure}
	\includegraphics[width=\columnwidth]{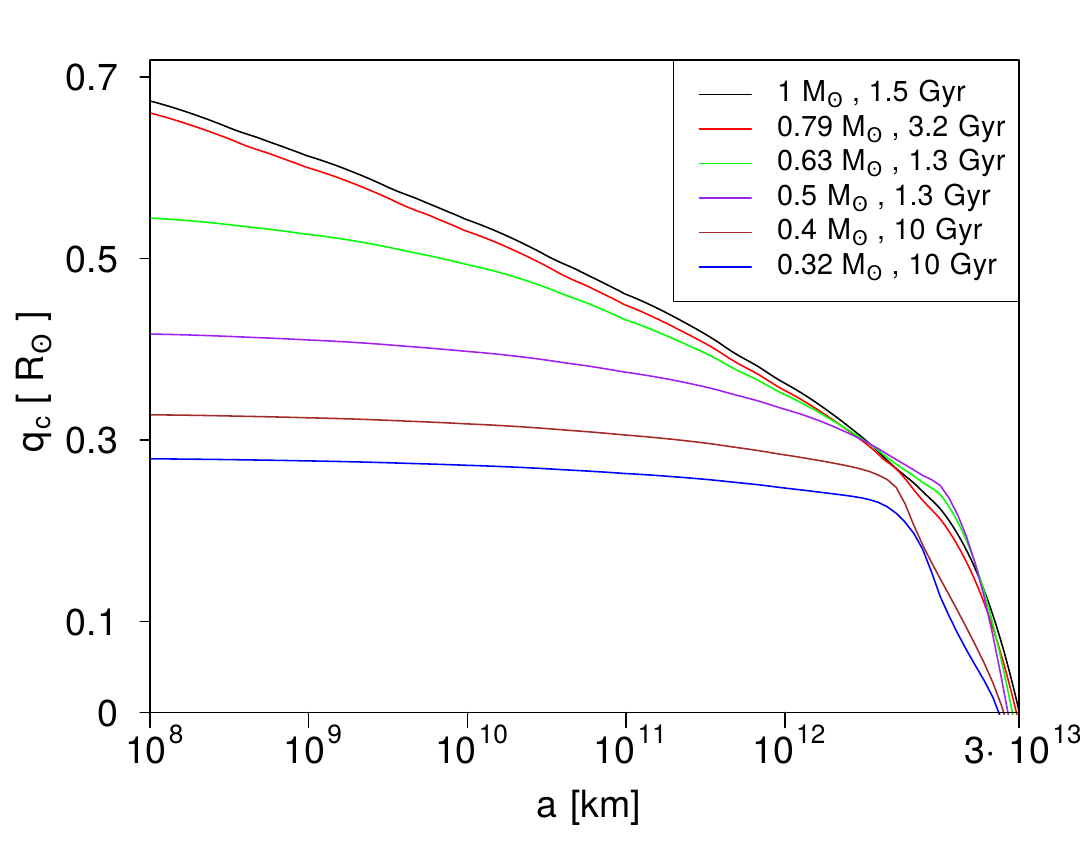}
    \caption{Critical extrapolated periastron $q_c$ of a PBH of $10^{-12} \msun$ as defined in section \ref{sec:PBH_num}, as a function of the initial semimajor axis $a$, for capture to occur in a time $t< 10^{10}$ years.
    }
    \label{fig:Critecc_1}
\end{figure}

Using our stellar models we compute the critical eccentricity as a function of the initial semimajor axis $a$ required for the PBH to be captured over a time $t_c= 10^{10}$ years, assuming there are no gravitational perturbations to the potential of the spherical star of mass $M_*$. All results shown from now on are for the fluid case, which is the valid one for stellar interiors,
with the use of equation (\ref{eq:Thun}), but use of equation (\ref{eq:Chandra}) results in only minor changes. Equation (\ref{eq:tc}) can be used for this purpose to a very good approximation, except when $a$ is not very large compared to $R_*$, but here we have carried out the exact calculation of the orbital evolution up to the point where the PBH orbit is completely absorbed in the stellar interior. The result is shown in Figure \ref{fig:Critecc_1}, as the critical extrapolated periastron $q_c(a)=a[1-e_c(a)]$ (or the periastron the orbit would have if the star were replaced by a point mass), for our six stellar models. Note that the probability for a random orbit of semimajor axis $a$ to have a periastron below $q_c$ is $2q_c/a$, if $q_c \ll a$.

As expected, $q_c$ has a very weak variation with $a$ when $a\ll a_m$, and then drops sharply when $a$ becomes close to $a_m$. 
The friction when the PBH moves through the dense stellar core determines the maximum semimajor axis $a_m$ where the energy loss allows the PBH to be captured in the time $t_c$. When the PBH starts at $a\ll a_m$, it can cross the star many times resulting in $q_c \sim R_{\ast}$, but if the crossing occurs in the outer envelope the friction is greatly reduced; this causes the slow decrease of $q_c$ with $a$. 

In the absence of orbital perturbations, PBHs that start on an orbit with eccentricity $e> e_c=1-q_c/a$ are the ones that are inside the loss-cylinder and will therefore be captured by the stars, and those outside the loss-cylinder will not be captured in the time $t_c$.

Finally, we list the results of our calculation for the mean extrapolated pericenter $\bar q_c$ in Table \ref{tab:XiforR}, for our six stellar models and several values of $a$ within which the average is made. As previously seen in figure \ref{fig:Critecc_1}, the average $q_c$ will be of the order of the stellar radius but as we get closer to $a_{m}$,  $\bar q_c$ falls as PBHs at such distances need to cross the core to get captured. If trying to reproduce the capture rate for a lower $a$ it is enough to use the corresponding value of $\bar q_c$ in the table on equation (\ref{eq:generalized_capture_rate}) with the new value for $a_m$.

\begin{table}
	\centering
	\setlength\tabcolsep{4pt}
	\begin{tabular}{lcccccr} 
    \multicolumn{6}{c}{}\\
    	\hline
		 $M_{\ast}\,  [\msun]$ & 1 & 0.79 & 0.63 & 0.50 & 0.40 & 0.32 \\
    
		 $R_{\ast}\,  [\rsun]$ & 0.91 & 0.76 & 0.57 & 0.43 & 0.33 & 0.28\\
		 
        $t\,  \rm [Gyr]  $ &  1.5 &  3.2  &  1.3 & 1.3  & 10 & 10 \\
    	\hline
    	 $a_{m}\,  [pc]$ &  0.98 &  0.93  &  0.88 &  0.83 & 0.78 & 0.72 \\
		 \hline
		 $a\,  [km]$ &   &  \  &  $\bar q_c\, (a) \,[\rsun]$ &   &  &  \\
		 \hline
        $5\cdot 10^8$& 0.64 &	0.63 &	0.52 &	0.40 &	0.31 &	0.27 \\
		\hline 
		 $10^9$& 0.64 &	0.62 &	0.52 &	0.40 & 0.31	 &	0.27 \\
		\hline
		 $3\cdot 10^9$& 0.63 &	0.62 &	0.52 &	0.40 &	0.31 &	0.27 \\
		\hline
		 $10^{10}$& 0.59 &	0.58 &	0.51 &	0.40 &	0.31 &	0.27\\
		\hline
		$3\cdot 10^{10}$&0.58 &	0.56 &	0.51 &	0.40 &	0.31 &	0.27 \\
		\hline
		 $10^{11}$& 0.53 &	0.51 &	0.48 &	0.39 &	0.31 &	0.27 \\
		\hline
		 $3\cdot 10^{11}$& 0.50 &	0.48 &	0.45 &	0.38 &	0.31 &	0.27 \\
		\hline
		 $10^{12}$& 0.44 &	0.43 &	0.41 &	0.36 &	0.30 &	0.26 \\
		\hline
		$3\cdot 10^{12}$& 0.38 & 0.38 &	0.36 &	0.33 &	0.28 &	0.25\\
		\hline
		$10^{13}$& 0.32 & 0.34 & 0.31 &	0.30 &	0.27 &	0.23 \\
		\hline
		$a_{m}$& 0.25 & 0.27 & 0.25 &	0.25 &	0.25 &	0.21 \\
		\hline
	\end{tabular}
	\caption{Values of  $\bar q_c$ for $10^{-12} \msun$ PBHs for various upper limits of $a$ and stellar models. $a_{m}$ is defined as the maximum a in which $e_c = 1$ results in capture within $t_c$, as discussed in section \ref{sec:Perturbations}, but we give $\bar q_c$ for various additional upper limits of $a$.}
	\label{tab:XiforR}
\end{table}

\subsection{Results for captured PBH}

\begin{figure}
	\includegraphics[width=\columnwidth]{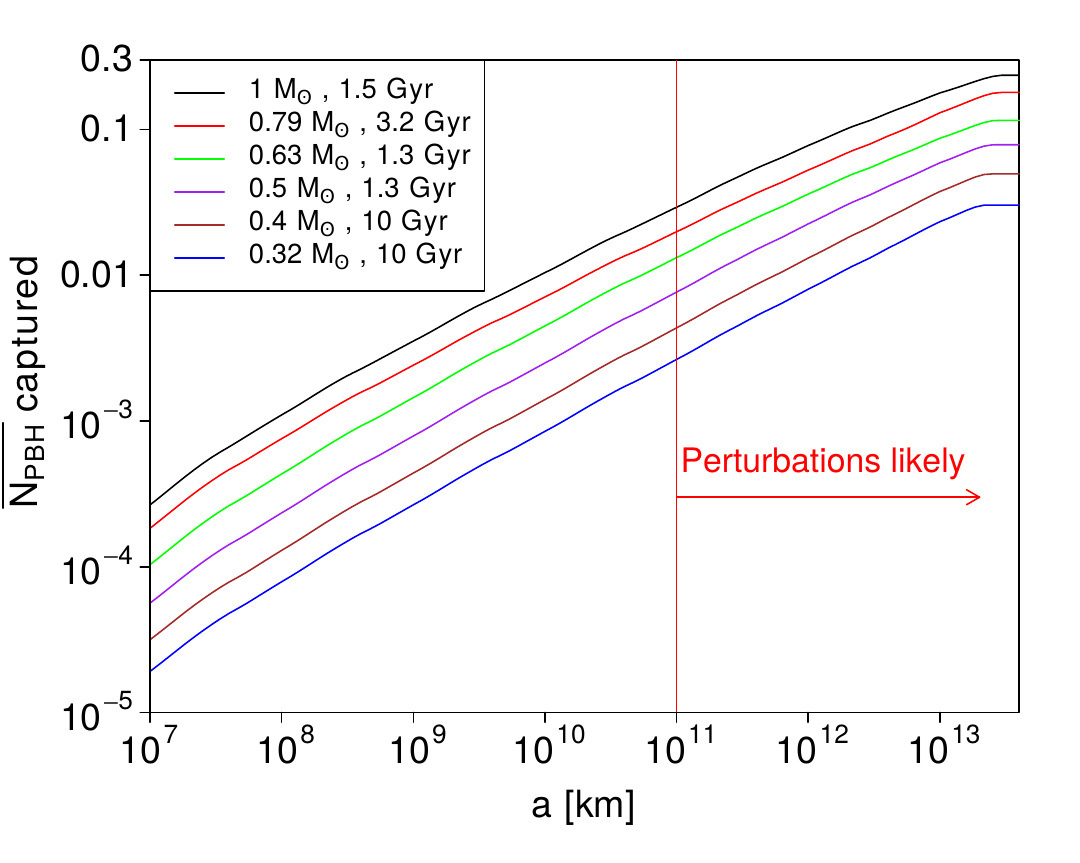}
    \caption{Mean number of PBHs captured by the star with initial semimajor axis within $a$, for the phase-space density of PBH at the halo radius $R=0.1 R_v$. The calculation is done for no external perturbations which we could expect to happen above the vertical red line but do not greatly alter the total capture rate, see main text for details.}
    \label{fig:PBH_capture_rate_all_stars}
\end{figure}

Figure \ref{fig:PBH_capture_rate_all_stars} shows the mean number of PBHs captured from orbits within an initial semimajor axis $a$. As expected from equation (\ref{eq:ncap}), this total number increases as $a^{1/2}$, except when $a$ is already close to $a_m$ and $q_c$ starts declining rapidly with $a$.

The assumption that there are no external tidal perturbations used for this calculation is not realistic, because even the tidal perturbation of the host DM halo becomes comparable to the acceleration by the star at $a\simeq 0.1$ pc, as discussed in section \ref{sec:Perturbations}. Even at the smaller radius $a\simeq 0.01$ pc (indicated by the vertical red line in the figure), where the tidal perturbation from the minihalo is $10^3$ times smaller than the stellar acceleration, the critical eccentricity is only $1-e_c=q_c/a\simeq 10^{-6}$, so the tidal perturbation can move the periastron outside the loss-cylinder in just one period, according to equation (\ref{eq:deg}).

However, while the perturbations eliminate any PBH captures from semimajor axes as large as $0.01$ pc, they should correspondingly increase captures from smaller $a$, as discussed in section \ref{sec:Perturbations}. As $a$ decreases, the loss-cylinder from which the PBH can be captured increases in width as $\delta e \propto a^{-3/2}$ if we assume that the PBH random-walks through the interval $\delta e$ owing to the tidal perturbations, spending a fraction of the total time $t_c$ proportional to $a^{1/2}$ in the region of width $\delta e_c\propto a^{-1}$ where the stellar core is crossed. This fraction of time is enough to capture a PBH that starts on an orbit at semimajor axis $a$. The eccentricity interval swept by the random-walk caused by an external tide narrows down as $a$ decreases, and when it coincides with the capture region with $\delta e\propto a^{-3/2}$, a total dominant capture rate is produced that is roughly independent of $a$ if the total number of PBH within $a$ increases as $a^{3/2}$.

Based on this argument, our calculation of the total capture rate shown in Figure \ref{fig:PBH_capture_rate_all_stars} should have a wide range of validity, even in cases where tidal perturbations are added from a variety of causes such as passing stars or crossing of galactic disks. The predicted total capture rate for a star that acquired its bound DM at birth at $R=0.1 R_v$, in our standard halo of $M_h=10^7\msun$ at $z=20$, is $\sim 0.3$ PBH for the $1\msun$ star, and 10 times lower for stars of $0.3\msun$. This is a substantial probability for low-mass stars born at this high-redshift to have formed low-mass stellar black holes by the present time. Tidal perturbations may still reduce this probability, for example if the bound DM is first tidally disrupted and then perturbations cease when the star is ejected to a region of very low density. But, as we have argued, our calculation should be realistic for many of the low-mass stars formed at high-redshift.

\begin{figure}
	\includegraphics[width=\columnwidth]{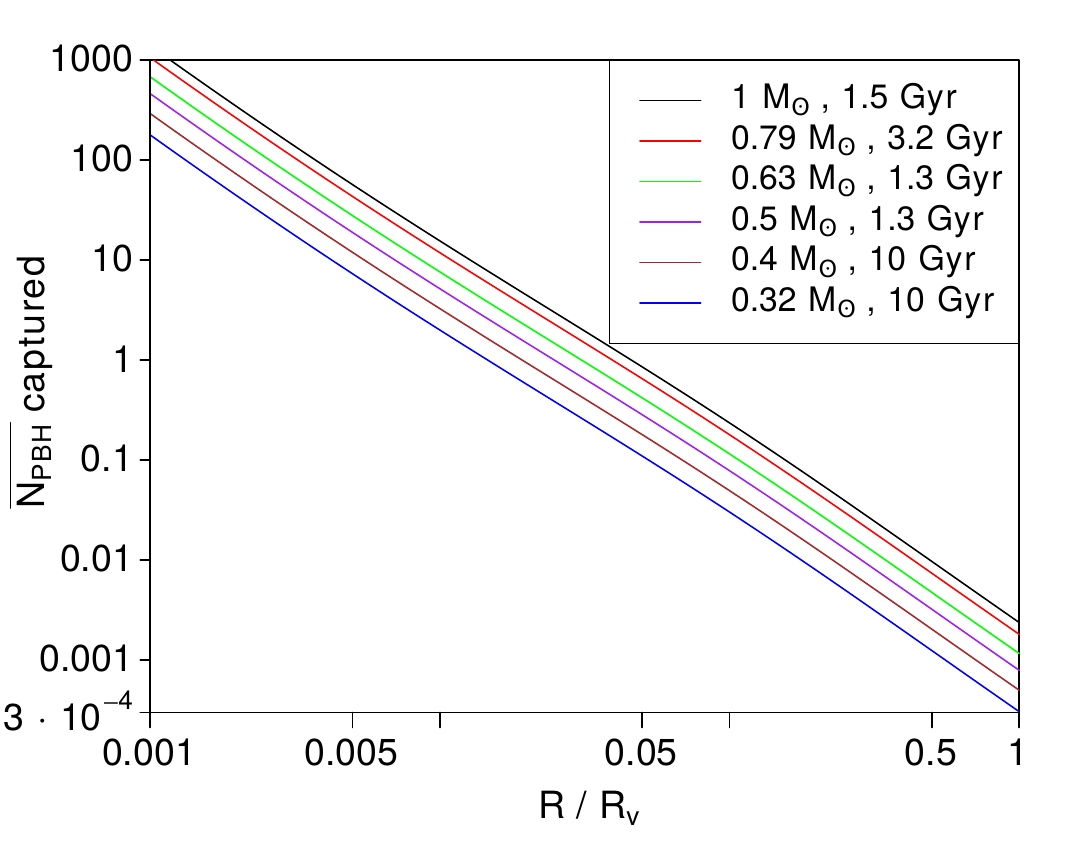}
    \caption{Mean number of PBHs captured by a star as a function of halo radius of the stellar birth site, for a NFW halo with DM halo with isotropic velocity dispersion.}
    \label{fig:PBH_capture_rate_Halo}
\end{figure}

Furthermore, many stars may form much closer to the halo center, increasing the phase-space density of the bound PBH acquired at birth. Figure \ref{fig:PBH_capture_rate_Halo} shows the result for the total capture rate, up to $a=a_m$, as a function of the initial halo radius $R$ where the star is formed, assuming the validity of our argument that tidal perturbations never reduce this rate. We assume the NFW profile with isotropic velocity dispersion, with the velocity dispersion profile shown in Figure \ref{fig:sigma_halo}, and the normalizing factor $Q$ for the bound DM density in equation (\ref{eq:rbd}), $Q \propto \rho_h/\sigma_h^3$. We see that the mean number of PBH captured actually reaches unity for stars born at $R\sim 0.03 R_v$, which is a typical formation radius for stars in present-day galaxies in galactic DM halos.

Considering PBH of different masses, we note that the specific energy loss in a stellar crossing is proportional to $m_b$, implying that lighter PBH are more difficult to capture, and that the maximum semimajor axis for capture $a_m$ increases as $m_b^2$ in equation (\ref{eq:ams}), so lighter PBH are captured only from smaller radii. However, for fixed $Q$ and therefore a fixed {\it mass} density of PBH, the number density is proportional to $m_b^{-1}$ and the total PBH capture rate is independent of $m_b$. We therefore conclude with the robust conclusion that if PBH over the broad asteroid-mass range account for most of the DM, a substantial fraction of low-mass stars formed at high redshift will inevitably form low-mass black holes after capturing a PBH and accreting onto them, and these low-mass stellar black holes should be present in the Universe today.

\section{Discussion and Conclusions} \label{sec:Conclusions}

Observational constraints for the abundance of PBHs have left only the asteroid-mass window, $10^{-16} \lesssim m_b/M_\odot \lesssim 10^{-11}$, as the one where all the DM may be composed of monochromatic PBHs \citep{2021RPPh...84k6902C}. Extended mass distributions of PBHs open additional windows \citep{2022ApJ...926..205C}, but constraining those distributions is a complex process. For an overview we refer to \citet{2017PhRvD..96b3514C, 2018JCAP...01..004B} . 

This manuscript shows that if the DM is indeed made of these asteroid mass PBHs, the first generation of low-mass stars formed at $z\sim 20$ in low-mass halos would have a high probability to make stellar black holes with masses less than a Chandrasekhar mass, through the process of capture of a PBH by the main-sequence star and the subsequent accretion and growth of the PBH. The final black hole may reach a mass comparable to the initial stellar mass, with uncertainties related to the possible formation of an accretion disk around the growing black hole. The reason is that the radiative efficiency of such a disc might be able to hamper accretion by dispersing the stellar material when only a small fraction of the stellar mass has been accreted. Other uncertainties include the possible role of external tidal perturbations or internal ones due to, for example, a planetary system around the star. The PBH capture rate would also be heavily modified in binary stars \citep{2012PhRvL.109f1301B}. Further work will be needed to clarify some of these uncertainties; nevertheless, the calculations presented in this paper suggest as a likely outcome of this asteroid-mass PBH scenario that many low-mass stellar black holes formed in this process may exist today in the Milky Way, after having originated in early dwarf galaxies that later merged into our Galaxy.

This paper extends earlier works \citep{2009ApJ...705..659A,2009MNRAS.399.1347B,2013PhRvD..87b3507C} with
calculations that use models of very metal-poor main-sequence stars of low mass, an analysis of the impact of external tidal perturbations on the PBH capture rate (with an improved treatment compared to previous work, e.g. \citet{2013PhRvD..87l3524C, 2014PhRvD..90h3507C}), and its dependence on the PBH mass. We reach the remarkable conclusion that the capture of PBH by these low-mass stars formed in the early DM halos with highest phase-space densities should occur for most stars formed within a halo radius $R\sim 0.03 R_v$ (a typical location for star formation in galactic halos), over this whole asteroid-mass range for PBH.

If many of these early low-mass stars have indeed collapsed to low-mass stellar black holes, their remnants should be found today among the Milky Way stellar populations, because the early dwarf galaxies where they formed may merge into more massive halos and end up tidally disrupted in the Milky Way halo over a wide range of radii. Their spatial distribution in the Milky Way is not easy to predict: early simulations proposed that
the remnants of the most ancient stellar populations should be found near the centre of the Galaxy, owing to the high bias factor of DM halos formed at high redshift \citep{2000fist.conf..327W,2006ApJ...653..285S}, but more recent work has found that baryonic effects may imply a broader radial distribution over the outer halo \citep{2018MNRAS.480..652E}.

How could these low-mass stellar black holes be detected at present? The most obvious possibility is via gravitational wave emission, if the black hole is in a binary system that leads to a merger. Some binary systems might contain two stars that both collapsed to low-mass black holes below the Chandrasekhar mass; alternatively, one of these low-mass black holes might be left in orbit around a more massive star that collapses to a black hole or neutron star following the standard evolution. In both cases, these binaries could spiral down into a merger at the present time that is detectable by the LIGO-Virgo collaboration \citep{2015CQGra..32g4001L,2021arXiv211103634T}. The detection of a merging black hole with a mass below the Chandrasekhar value would naturally indicate new physics beyond standard stellar evolution theory, and has already been searched in recent gravitational wave searches \citep{2021arXiv210912197T}.

Another way of detecting these low-mass black holes is when they are observed in a binary system where the other object is a luminous star. For a main-sequence companion, a normal star would be seen orbiting around a dark object. The difficulty in this case would be to rule out an old white dwarf as the unseen companion, because white dwarfs are very faint and difficult to see when they are unresolved from the main-sequence star. If the companion is a white dwarf then the system is much fainter and difficult to discover, but it is then easier to rule out another white dwarf companion for an unseen compact object below the Chandrasekhar mass. The additional difficulty to accomplish this type of detection is that the low-mass black holes formed from PBH capture would be rare and present only among very low-metallicity stars, so a lot of these possible binaries would have to be examined, most of which would contain regular old white dwarf companions.

If the low-mass black hole that is formed by a PBH captured by a star is isolated, then it is extremely difficult to discover. Microlensing seems the only possibility \citep{1986ApJ...304....1P, 2000ApJ...542..281A}, but these black holes would actually pass for M-dwarfs, white dwarfs or brown dwarfs, depending on their mass. These other objects are all very faint and therefore usually not possible to distinguish from low-mass black holes, with a much lower expected abundance in our PBH scenario.

In summary, the open asteroid-mass window for PBH as DM is a possibility in which we expect that low-mass black holes of stellar mass, but below the Chandrasekhar limit, exist today. These can be found if they are in binaries that are either tight enough to lead to mergers with other compact objects detectable via gravitational waves at present, or through the direct detection of the binary companion in the Milky Way and identification of the unseen object as a low-mass black hole. These binaries are expected to be among the first stellar systems to have formed, and therefore of very low-metallicity, which may help in their identification among halo stars in our vicinity or closer to the Milky Way centre.

\section*{Acknowledgements}


We would like to acknowledge helpful discussions and advice from  N. Bellomo, J. L. Bernal, A. Escriv\`a, C. Germani, and J. Salvad{\'o}. This work was supported in part by Spanish grants CEX-2019-000918-M funded by MCIN/AEI/10.13039/501100011033, AYA2015-71091-P, and PID2019-108122GB-C32.

\section*{Data availability}

All data underlying this article not shown will be shared on reasonable request to the first author.




\bibliographystyle{mnras}
\typeout{}
\bibliography{PBH_capture}{} 




\appendix


\bsp	
\label{lastpage}
\end{document}